\documentclass[aps,prd,onecolumn,showpacs,nofootinbib,groupedaddress]{revtex4}  
\usepackage{graphicx}
\usepackage{epstopdf}
\usepackage{amsmath}
\usepackage{amsfonts}
\usepackage{amssymb}
\usepackage{appendix}
\usepackage{comment}
\usepackage{bbold}
\usepackage{color}
\usepackage{slashed}
\usepackage{subfigure}
\usepackage{setspace}
\usepackage{footnote}
\usepackage{multirow}
\usepackage{longtable}
\usepackage{braket}

\def\ra{\rightarrow}
\newcommand{\optbar}[1]{\shortstack{{\tiny (\rule[.4ex]{1em}{.1mm})} 
  \\ [-.7ex] $#1$}}		
\newcommand{\Eq}[1]{Eq.~(\ref{eq#1})}
\newcommand{\beq}{\begin{equation}}
\newcommand{\eeq}{\end{equation}}

\begin{document}

\singlespacing

{\hfill FERMILAB-PUB-18-418-T, NUHEP-TH/18-09}

\title{Addressing the Majorana vs. Dirac Question with Neutrino Decays}

\author{A. Baha Balantekin}
\affiliation{Department of Physics, University of Wisconsin, Madison, WI 53706, USA}
\author{Andr\'{e} de Gouv\^{e}a} 
\affiliation{Northwestern University, Department of Physics \& Astronomy, 2145 Sheridan Road, Evanston, IL 60208, USA}
\author{Boris Kayser}
\affiliation{Theoretical Physics Department, Fermilab, P.O. Box 500, Batavia, IL 60510, USA}

\begin{abstract}
The Majorana versus Dirac nature of neutrinos remains an open question. This is due, in part, to the fact that virtually all the experimentally accessible neutrinos are ultra-relativistic.
Noting that Majorana neutrinos can behave quite differently from Dirac ones when they are non-relativistic, we show that, at leading order, the angular distribution of the daughters in the decay of a heavy neutrino into a lighter one and a self-conjugate boson is isotropic in the parent's rest frame if the neutrinos are Majorana fermions, independent of the parent's polarization. This result follows from CPT invariance and is independent of the details of the physics responsible for the decay. In contrast, if the neutrinos are Dirac fermions, the angular distribution in such a decay is, in general, not isotropic.
 We explore the feasibility of using these angular distributions---or, equivalently, the energy distributions of the daughters in the laboratory frame---in order to address the Majorana versus Dirac nature of neutrinos if a fourth, heavier neutrino mass eigenstate reveals itself in the current or next-generation of high-energy colliders, intense meson facilities, or neutrino beam experiments. We also point out how the related decays of a heavy neutrino into charged daughters can be used for the same purpose.
\end{abstract}


\maketitle

\setcounter{equation}{0}
\setcounter{footnote}{0}
\section{Introduction}
\label{sec:introduction}

One of the leading unanswered questions about the neutrinos is whether they are Majorana or Dirac particles. Since all neutrinos that have been directly observed so far have been ultra-relativistic in the rest frame of the observing experiment, and ultra-relativistic Majorana neutrinos will almost always behave just like Dirac ones, the effort to determine whether neutrinos are Majorana or Dirac particles has proved very challenging. The most promising approach, by far, that is presently being pursued is the search for neutrinoless double beta decay. 

In contrast to the behavior of ultra-relativistic neutrinos, that of {\em non-relativistic} ones can depend quite a lot on whether they are of Majorana or Dirac character. This is illustrated by the capture rate on tritium of the relic neutrinos from the Big Bang. Many, and perhaps all, of these very cold neutrinos are non-relativistic. For a given density, the tritium capture rate of the non-relativistic ones is twice as large if they are Majorana particles as it is if they are Dirac particles \cite{Long:2014zva}. Unfortunately, the capture rate also depends on other unknowns, including the actual local (not universe-average) relic neutrino density, so using tritium capture of the relic neutrinos to determine whether neutrinos are of Majorana or Dirac character may prove to be unfeasible. Very low-energy $e\gamma\to e\nu\bar{\nu}$ scattering \cite{Berryman:2018qxn} and neutrino pair emission from excited atoms \cite{Yoshimura:2011ri,Yoshimura:2006nd,Fukumi:2012rn,Dinh:2012qb,Song:2015xaa} have also been explored as sources of non-relativistic neutrinos capable of addressing the Majorana versus Dirac question. The rates for these and related processes, alas, are exceedingly small. 

The observation that non-relativistic Majorana and Dirac neutrinos can behave quite differently leads us to consider the possibility that there is a heavy neutrino $N$ whose decays could be studied. In its rest frame, this neutrino would obviously be totally non-relativistic. Such a neutrino is being sought experimentally (for recent experimental efforts, see, for example, \cite{Aad:2015xaa,Sirunyan:2018xiv,Cenci:2018nlo,Anelli:2015pba,Aguilar-Arevalo:2017vlf}. Recent compilations of existing experimental searches and constraints can be found in, for example, Refs.~\cite{Smirnov:2006bu,Atre:2009rg,Drewes:2015iva,deGouvea:2015euy,Alekhin:2015byh,Fernandez-Martinez:2016lgt,Adhikari:2016bei,Abada:2016plb,Cai:2017mow,Drewes:2018gkc,Deppisch:2015qwa}.). Given that the leptons are known to mix, if one neutrino mass eigenstate is a Majorana fermion, it is very likely that all the mass eigenstates, including $N$, are Majorana fermions. Consequently, in this work we assume that either all neutrino mass eigenstates are Majorana fermions, or else all of them are Dirac fermions.

If all neutrinos are Majorana fermions, the rate for $N$ to decay into some specific final states is twice as large as it would be if all neutrinos were Dirac particles \cite{Gorbunov:2007ak}. However, this difference may not be too useful because the rate for decay into a given final state also depends on unknown parameters: active--sterile mixing angles, the existence of other new particles and interactions, etc. Thus, it is intriguing that the Majorana or Dirac character of neutrinos could also be revealed by the angular distribution of the particle $X$ in a decay of the form $N\rightarrow \nu_l + X$, where $\nu_l$ is a lighter neutrino and $X$ is a self-conjugate boson. The angular distributions in decays of this kind, and the laboratory-frame energy distributions of particle $X$ that correspond to them, are the focus of this paper. The angular distributions in the related decays $N \ra \ell_\alpha^\mp + X^\pm$, where $\ell_\alpha$ is a charged lepton, can also be revealing, and will be discussed as well.

If there is a heavy neutrino $N$, the observation at, for example, a hadron collider of a lepton-number nonconserving sequence such as $\mathrm{quark + antiquark} \rightarrow W^+ \rightarrow N + \mu^+ \rightarrow (e^+ \pi^-) + \mu^+$ would tell us that the neutrinos, including $N$, are Majorana particles \cite{Keung:1983uu}. However, this type of information is not always experimentally available. For example, if $N$ is discovered at a neutrino oscillation experiment, manifest lepton-number nonconservation involving like-sign leptons as in our illustrative sequence may be impossible to establish because the detector may not have charge discrimination. The angular distributions on which we focus here could nonetheless still be studied.

\setcounter{equation}{0}
\setcounter{footnote}{0}
\section{Neutrino Decay}
\label{sec:decay}

Here we consider the two-body decays $N \ra \nu_l + X$ of a heavy, polarized, spin one-half, neutral fermion mass eigenstate $N$.\footnote{In what follows, nothing precludes $N$ from being one of the established light neutrino mass eigenstates $\nu_1,\, \nu_2$, or $\nu_3$. In this case, in the absence of new, very light particles, the only accessible two-body decay is $N \ra \nu_l + \gamma$.}  A preliminary version of the following discussion was given in Refs.~\cite{Balantekin:2018azf,Kayser:2018aa}. The daughter fermion $\nu_l$ is a lighter neutral fermion, possibly one of the established light neutrino mass eigenstates $\nu_1,\, \nu_2$, or $\nu_3$, and $X$ is a self-conjugate boson. Depending on the mass of $N,\; X$ could, for example, be a $\gamma,\, \pi^0,\, \rho^0, \,Z^0$, or $H^0$. If $X$ is any of these particles, the decay rate $\Gamma(N \ra \nu_l + X)$ is twice as large if $N$ and $\nu_l$ are Majorana particles as it is if they are Dirac particles \cite{Gorbunov:2007ak}. However, as already noted, this difference may not be a useful way to tell whether neutrinos are Majorana or Dirac particles, because the decay rate $\Gamma(N \ra \nu_l + X)$ also depends on other unknown parameters. Thus, it is fortunate that the angular distribution of the daughters, which in most of these decay modes does not depend on elusive unknown parameters, is also quite sensitive to whether neutrinos are Majorana or Dirac particles.

\subsection{Decay properties}

\begin{figure}[h]
\includegraphics[width=3in]{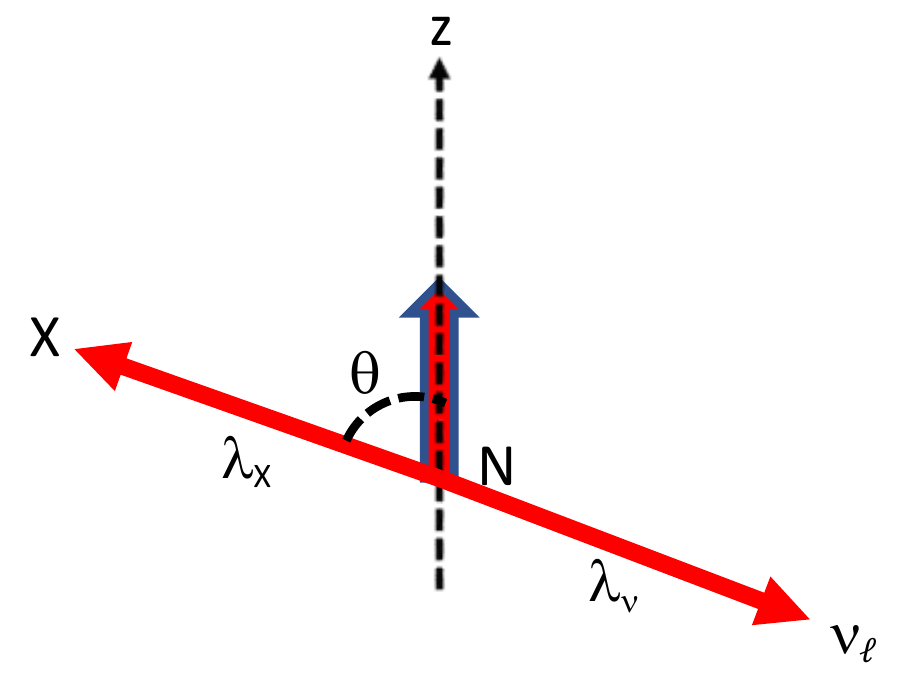}
\caption{The decay $N \ra \nu_l + X$.}
\label{decayfig}
\end{figure}

Let us consider, in the parent's rest frame, the decay $N \ra \nu_l + X$ of a heavy neutrino $N$ that is fully polarized by its production mechanism, with its spin pointing along a direction we shall call $+z$. Suppose that the particle $X$ emerges at an angle $\theta$ with respect to the $+z$ direction (with $\nu_l$ emerging oppositely), and that $X$ and $\nu_l$ are produced with helicities $\lambda_X$, and $\lambda_\nu$, respectively (see Figure \ref{decayfig}). 
With $\lambda \equiv \lambda_X - \lambda_\nu$, rotational invariance dictates that the angular distribution of $X$ is given by
\beq
\frac{d\Gamma (N \ra \nu_l + X)}{d(\cos \theta)} = \frac{\Gamma_{\lambda = + 1/2}}{2} (1 + \cos \theta) +  \frac{\Gamma_{\lambda = - 1/2}}{2} (1 - \cos \theta) \;\; .
\label{eq1}
\eeq
Here, $\Gamma_{\lambda = + 1/2}$ is the total rate for decays $N \ra \nu_l + X$ yielding daughter helicity configurations that have $\lambda = + 1/2$, and similarly for $\Gamma_{\lambda = - 1/2}$. We may rewrite the angular distribution of \Eq{1} as
\beq
\frac{d\Gamma (N \ra \nu_l + X)}{d(\cos \theta)} = \frac{\Gamma}{2} (1 + \alpha \cos\theta) \;\; ,
\label{eq2}
\eeq
where 
\beq
\Gamma =  \Gamma_{\lambda = + 1/2} + \Gamma_{\lambda = - 1/2} > 0 \;\; ,
\label{eq3}
\eeq
and
\beq
\alpha = (\Gamma_{\lambda = + 1/2} - \Gamma_{\lambda = - 1/2}) / \Gamma  \;\; \in [-1,+1] 
\label{eq4}
\eeq
is the asymmetry parameter.

For the moment, let us suppose that neutrinos are Dirac particles, and that the decays described by Eqs.~(\ref{eq1}-\ref{eq4}) are those of {\em neutrinos}. For the antineutrino decays, we have, in analogy to Eqs.~(\ref{eq1}) and (\ref{eq2}),
\begin{eqnarray}
\frac{d\Gamma (\bar{N} \ra \bar{\nu}_l + X)}{d(\cos \theta)} &=& \frac{\bar{\Gamma}_{\lambda = + 1/2}}{2} (1 + \cos \theta) +  \frac{\bar{\Gamma}_{\lambda = - 1/2}}{2} (1 - \cos \theta) \nonumber \\ 
	&=& \frac{\bar{\Gamma}}{2} (1 + \bar{\alpha} \cos\theta) \;\; ,
\label{eq7}
\end{eqnarray}
where the parameters $\bar{\Gamma}_{\lambda = + 1/2},\; \bar{\Gamma}_{\lambda = - 1/2}, \; \bar{\Gamma}$, and $\bar{\alpha}$ are the $\bar{N}$ decay analogues of their $N$ decay counterparts.

At leading order in perturbation theory, the $N$ decay amplitude for given $\theta$ and daughter helicities is \beq
\langle X(\theta, \lambda_X)\, \nu_l (\pi-\theta,\lambda_\nu) \;|\; {\cal H}_{\mathrm{int}}\; |\; N(\mathrm{up}) \rangle \;\; .
\label{eq5}
\eeq
Here, ${\cal H}_{\mathrm{int}}$ is the Hamiltonian, or effective Hamiltonian, that causes the decay, and the ``up'' indicates that the parent $N$ spin points in the $+z$ direction. We assume  that ${\cal H}_{\mathrm{int}}$ is invariant under CPT $\equiv \zeta: \; \zeta {{\cal H}_{\mathrm{int}} }\zeta^{-1}= {\cal H}_{\mathrm{int}}$. Then, taking into account that CPT is an antiunitary operator, 
\begin{eqnarray}
 | \langle X(\theta, \lambda_X) \,\nu_l (\pi-\theta,\lambda_\nu) \;|\; {{{\cal H}_{\mathrm{int}}}}\; |\; N(\mathrm{up}) \rangle | ^2   &=&
  |\;\langle \zeta {{\cal H}_{\mathrm{int}} } \zeta^{-1}\zeta N(\mathrm{up}) \;|\; \zeta X(\theta, \lambda_X) \, \nu_l (\pi-\theta,\lambda_\nu)\rangle \;|^2   \nonumber \\
   &=& |\; \langle {{\cal H}_{\mathrm{int}} } \bar{N}(\mathrm{down}) \;|\; X(\theta, -\lambda_X) \, \bar{\nu}_l(\pi-\theta,-\lambda_\nu)\rangle \;|^2   \nonumber \\
  &=& |\;  \langle X(\pi-\theta, -\lambda_X) \,\bar{\nu}_l(\theta,-\lambda_\nu) \;|\;  {\cal H}_{\mathrm{int}}\; |\;  \bar{N}(\mathrm{up}) \rangle \;|^2 \;\; .
\label{eq6}
\end{eqnarray}
Here, the last step assumes invariance under a 180$^\circ$ rotation about the axis perpendicular to the decay plane. 

Owing to the antiunitarity and antilinearity of $\zeta$, the CPT invariance of ${\cal H}_{\mathrm{int}}, \; \zeta{\cal H}_{\mathrm{int}} \zeta^{-1}= {\cal H}_{\mathrm{int}}$, does not imply that the all-orders transition operator ${\cal T}$ for $N \ra \nu_l + X$ obeys $\zeta{\cal T}\zeta^{-1} = {\cal T}$, but only that it obeys $\zeta {\cal T}\zeta^{-1} = {\cal T}^\dagger$. For this reason, the constraint of \Eq{6} holds only in lowest order, where ${\cal T} = {\cal H}_{\mathrm{int}}$, a Hermitean operator for which ${\cal H}_{\mathrm{int}}^\dagger = {\cal H}_{\mathrm{int}}$.  Henceforth, unless otherwise noted, we assume that the lowest order result is an excellent approximation for the full result. 

Summed over the helicities for which $\lambda_X - \lambda_\nu \equiv \lambda = + 1/2$, \Eq{6} implies that $\Gamma_{\lambda = + 1/2} = \bar{\Gamma}_{\lambda = - 1/2}$. Similarly, summed over the helicities for which $\lambda = -1/2$, it implies that $\Gamma_{\lambda = - 1/2} = \bar{\Gamma}_{\lambda = + 1/2}$. It follows that
\beq
\bar{\Gamma} = \Gamma \;\; ,
\label{eq8}
\eeq
and that
\beq
\bar{\alpha} = -\alpha \;\; .
\label{eq9}
\eeq

Now, suppose that neutrinos are not Dirac particles, but Majorana ones. \Eq{6} still holds, but with the bars distinguishing antineutrinos from neutrinos erased. The neutrino decay angular distribution is described by Eqs.~(\ref{eq1}-\ref{eq4}) and now \Eq{6}, summed over the helicities for which $\lambda = + 1/2$, implies that $\Gamma_{\lambda = + 1/2} = \Gamma_{\lambda = - 1/2}$. That is,
\beq
\alpha = 0 \;\; ;
\label{eq10}
\eeq
the angular distribution is isotropic in the case of Majorana neutrino decay. This isotropy was noted for the special case where $X=\gamma$ in Refs.~\cite{Li:1981um, Shrock:1982sc}. As we see, it holds for any self-conjugate boson $X$. As we also see, it is a consequence of rotational and CPT invariance alone, and does not depend on any further details of the interactions(s) driving the decay.\footnote {We thank S. Petcov for long-ago discussions of this point for the case where $X=\gamma$.}

If the transition operator ${\cal T}$ for the decay $N \ra \nu_l + X$ is CP invariant, then 
\beq
 | \langle X(\theta, \lambda_X) \,\nu_l (\pi-\theta,\lambda_\nu) \;|\; {\cal T}\; |\; N(\mathrm{up}) \rangle | ^2  = | \langle X(\pi-\theta, -\lambda_X) \,\bar{\nu}_l(\theta,-\lambda_\nu) \;|\;  {\cal T} \; |\;  \bar{N}(\mathrm{up}) \rangle \;|^2 \;\; .
\label{eq11}
\eeq
This is the same constraint that we obtained from CPT invariance, but since CP, unlike CPT, is a unitary operator, there is no longer any requirement that ${\cal T}$ be Hermitean so, if CP invariance holds, the constraint holds to all orders in perturbation theory. Of course, the transition operator ${\cal T}$ may very well violate CP invariance. If it does, then the constraint of \Eq{11} is invalid, but the CPT constraint of \Eq{6} on the lowest-order decay amplitude still holds. For the processes we are considering, the lowest-order amplitude is likely to be an excellent approximation.
 
In contrast to its isotropy in the Majorana case, the angular distribution in $N \ra \nu_l + X$ need not be isotropic in the Dirac case. Indeed, as we discuss in Section~\ref{sec:application}, if one assumes the decay of the neutrino is governed by the Standard Model weak interactions, the angular distributions of the various neutrino decay modes are typically quite far from isotropic and might allow us to determine whether neutrinos are Majorana or Dirac particles.

\subsection{Energy distribution in the Laboratory}

In the previous subsection, we considered the angular distribution of the decay of a neutral fermion in its rest frame. Given that the daughter $\nu_{l}$ from the $N \rightarrow \nu_{l} +X$ decay is likely to fly off the detector environment undetected, reconstructing the $N$ rest frame on an event-by-event basis may prove to be, experimentally, very challenging.\footnote{If the four-momentum of the $X$ particle were measured and if the direction of the momentum of the parent particle were known, it would be possible to reconstruct the four-momentum of $N$ on an event-by-event basis.} Here, instead, we consider the decays in the laboratory frame and consider the energy distribution of the $X$ particle, which ``inherits'' the properties of the angular distribution of the $X$ particle in the rest frame of the neutral heavy lepton.   

If the neutral heavy lepton has a fixed laboratory energy $E_N$, and correspondingly a fixed laboratory three-momentum of magnitude $p_N$, the $X$ particle is produced in the decay $N \rightarrow \nu_{l} +X$ with laboratory energies $E_X^{(L)}$ that range from
\begin{equation}
\label{maximumEX}
E_X^{\rm (L, min)} = \frac{1}{2} \left[ E_N \left( 1 + r \right) - p_N \left( 1 - r \right) \right] ,
\end{equation}
to
\begin{equation}
\label{minimumEX}
E_X^{\rm (L, max)} = \frac{1}{2} \left[ E_N \left( 1 + r \right) + p_N \left( 1 - r \right) \right] ,
\end{equation}
assuming the daughter $\nu_{l}$ to be massless. Here, $r=m_X^2/m_N^2<1$, $m_N$ is the mass of the parent neutrino $N$, and $m_X$ is the mass of the daughter boson $X$. 

If the $X$ particle has the angular distribution, in the parent's rest frame,
\begin{equation}
\frac{dn_X}{d \cos \theta_X} \propto ( 1+ A  \cos \theta_X),
\label{eq:rest}
\end{equation}
where $A\equiv \alpha P$, $\alpha$ is the decay asymmetry parameter introduced in the last subsection, and $P$ is the polarization of the $N$ sample, it is straightforward to compute the energy distribution of $X$ in the laboratory frame:
\begin{equation}
\frac{d n_X(E_N, E_X^{(L)})}{d E_X^{(L)}} \propto \frac{2}{p_N (1-r)}\left[ 1 + A \left( \frac{2}{(1-r)}\frac{E_X^{(L)}}{p_N}- \left( \frac{1+r}{1-r} \right) \frac{E_N}{p_N} \right) \right] .
\label{eq:lab}
\end{equation}
The energy distribution in the laboratory frame is linear in $ E_X^{(L)}$ and the slope of the distribution is proportional to $A$. Positive (negative) $A$ implies a harder (softer) energy distribution for $X$ in the lab frame. 

If, in an experimental setup,  the $N$ particles enter the detector as a beam with energy distribution $\rho(E_N)$,  the number of decay $X$ particles with energy  $E_X^{(L)}$ observed inside the detector with total length $\ell_D$ is proportional to 
\begin{equation}
\ell_D \int_{E_N^{\rm (min)}}^{E_N^{\rm (max)}} dE_N \frac{m_N}{p_N} \rho(E_N) \left[ \frac{d n_X(E_N, E_X^{(L)})}{d E_X^{(L)}} \right] \; .
\label{eq16}
\end{equation}
Here we assume the decay length of $N$ to be much longer than $\ell_D$. The integration limits are 

\begin{equation}
2 r E_N^{\rm (max,min)} = E_X^{(L)} (1+r) \pm (1-r) \sqrt{\left( \left(E_X^{(L)} \right)^2 - m_X^2 \right)} \;\; ,
\label{eq17}
\end{equation}
where the plus (minus) sign gives the maximum (minimum) value.

\begin{figure}[h]
\includegraphics[width=0.6\textwidth]{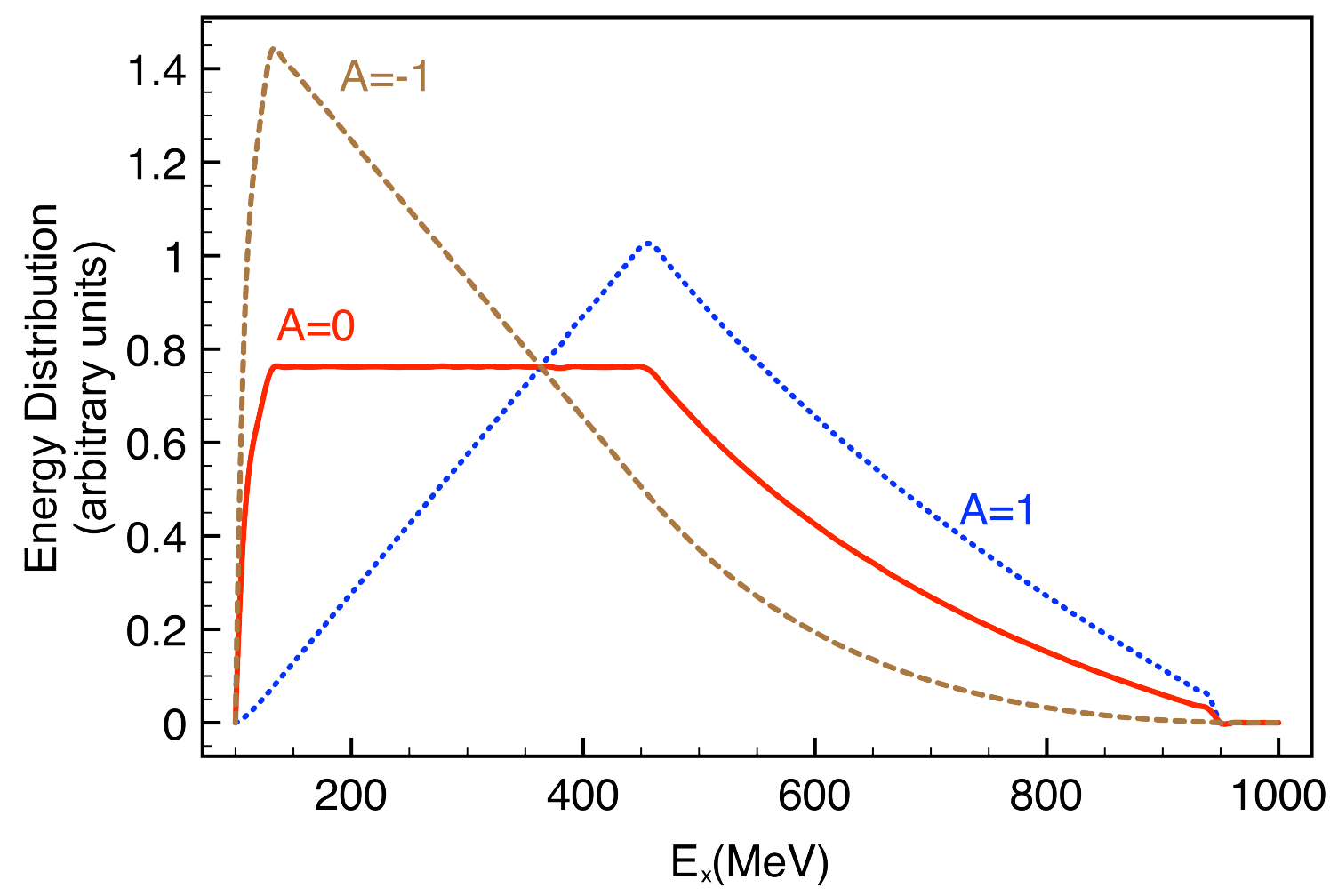}
\caption{Laboratory energy distributions of the daughter $X$ boson, assuming $m_X =$ 100 MeV, $m_N$ = 300 MeV, and a flat $E_N$ distribution bounded by 500 MeV and 1000 MeV, for $A=0,\pm 1$, defined in Eq.~(\ref{eq:rest}). If $N$ is a Majorana fermion, only $A=0$ is allowed, while any $A\in[-1,1]$ is possible if $A$ is a Dirac fermion.}
\label{MajvdDir}
\end{figure}

Figure~\ref{MajvdDir} depicts the rate of $X$ particles per unit energy as a function of the energy $E_X^{(L)}$, for  $m_X =$ 100 MeV, $m_N$ = 300 MeV, and a flat $E_N$ distribution bounded by 500 MeV and 1000 MeV. The different curves correspond to $A=0,\pm 1$. If $N$ is a Majorana fermion, only $A=0$ is allowed, while any $A\in[-1,1]$ is possible if $N$ is a Dirac fermion. At least in this case, the three curves are quite distinct and, naively, it seems that distinguishing Dirac from Majorana neutrinos using this energy distribution is straightforward as long as $|A|$ is not too small in the Dirac case.
 
The shapes of the curves in Figure~\ref{MajvdDir} are easy to understand. For 132 MeV $<  E_X^{(L)}  <$ 456 MeV, the entire nonzero spectrum of $N$ particles from 500 MeV to 1000 MeV can contribute to the event rate. That is, for $E_X^{(L)}$ in this range, the effective limits of integration in Eq.~(\ref{eq16}) do not depend on $E_X^{(L)}$. Moreover, for $A=0$, Eq.~(\ref{eq:lab}) shows that $dn_X  / d E_X^{(L)}$ does not depend on $E_X^{(L)}$ either. That is why the $A=0$ curve in Fig.~\ref{MajvdDir} is flat for 132 MeV $<  E_X^{(L)} <$~456~MeV. From the $E_X^{(L)}$ dependence of $dn_X  / d E_X^{(L)}$, Eq.~(\ref{eq:lab}), we see that the $A = -1$ curve in Fig.~\ref{MajvdDir} should have a negative slope for $E_X^{(L)} >$ 132 MeV and the $A = +1$ curve should have a positive slope for $E_X^{(L)} <$ 456 MeV. The three curves meet at one point, $E_X^{(L)}\sim 360$~MeV. This is a consequence of the fact that in $dn_X  / d E_X^{(L)}$, Eq.~(\ref{eq:lab}), the coefficient of $A$ contains two contributions of opposite sign, one of which depends on $E_X^{(L)}$. As a result, when the integral of \Eq{16} is performed, the term proportional to $A$ vanishes at the point $E_X^{(L)} \sim$ 360 MeV. At this point, the event rate is independent of $A$. For larger (smaller) $E_X^{(L)}$ values, the event rate for positive (negative) values of $A$ exceeds that for $A=0$.

\setcounter{equation}{0}
\setcounter{footnote}{0}
\section{Application---Neutral Heavy Leptons}
\label{sec:application}

Neutral heavy leptons, sometimes referred to as sterile neutrinos and, when appropriate, right-handed neutrinos, are benign, well-motivated additions to the Standard Model. They are a natural side-effect of different mechanisms, including the renowned seesaw mechanism, that lead to nonzero neutrino masses.  They also serve as a possible solution to the so-called short-baseline neutrino anomalies, and are an excellent warm dark matter candidate that is consistent with the observation of the currently unaccounted-for astrophysical 3.5~keV X-ray line. For recent comprehensive reviews on neutral heavy leptons, see Refs.~\cite{Abazajian:2012ys,Adhikari:2016bei,Abazajian:2017tcc}.

Neutral heavy leptons are the subject of intense experimental pursuit. Non-observations translate into constraints on their properties, especially their masses and how much they mix with the Standard Model (active) neutrinos. The simplest recipe for neutral heavy leptons, the one we will consider here unless otherwise noted, is as follows. Add to the Standard Model field content gauge-singlet fermions. After spontaneous symmetry breaking, these mix with the active neutrinos in such a way that the number of neutrino mass eigenstates is $n$, an integer larger than three. As usual, the flavor and mass eigenstates are related by a unitary matrix $U$ with elements $U_{\alpha i}$: $\nu_{\alpha}=U_{\alpha i}\nu_i$, where $\alpha=e,\mu,\tau,s_1,s_2,\ldots$, with $s$ labeling the new fermions, and $i=1,2,3,\ldots, n$ labeling the neutrino mass eigenstates, whose  masses are $m_{1,2,3,\ldots,n}$, respectively. We will assume that the neutrino masses are ordered from smallest to largest. The neutral heavy leptons, or heavy neutrinos, are $\nu_4, \nu_5, \ldots, \nu_n$. Unless otherwise noted, in this section we will refer to the heavy neutrinos generically as $\nu_4$, rather than as $N$ as in the previous sections, since ``$\nu_4$'' is a more natural notation for our present purpose.

Since the new gauge-singlet fermions do not couple to the $Z$-boson or the $W$-boson, the weak currents of the neutrino mass eigenstates are proportional to 
\begin{equation}
U_{\alpha i}\bar{\ell}_{\alpha}\gamma_{\mu}(1-\gamma_5)\nu_i~{\rm (charged~current)},
\end{equation}
where $\ell_{\alpha}$ are charged leptons, $\alpha=e,\mu,\tau$, or
\begin{equation}
\sum_{\alpha=e,\mu,\tau}U^*_{\alpha i}U_{\alpha j}\bar{\nu}_i\gamma_{\mu}(1-\gamma_5)\nu_j~{\rm (neutral~current)}.
\end{equation}
Assuming no new interactions, the production and decay of heavy neutrinos is described by the weak interactions and calculable as a function of the heavy neutrino masses and the elements of the mixing matrix. Depending on the heavy neutrino mass, heavy neutrinos are best probed by different experiments. For $m_4\lesssim 10$~eV, heavy neutrinos can be spotted in neutrino oscillation experiments with intense beams. For $m_4\lesssim 1$~GeV, heavy neutrinos are produced in the decay of charged and neutral mesons and can be looked for in intense meson facilities, including charm and $B$-factories. The existence of heavier neutrinos, $m_4\gtrsim 10$~GeV, can be effectively investigated in collider experiments. 

The heavy neutrino lifetime and the allowed neutrino decay modes also depend on the heavy neutrino mass. For masses below an MeV, only $\nu_4\to\nu_l\nu'_l\nu_l''$ and $\nu_4\to\nu_l\gamma$, $l=1,2,3$ are allowed. Above an MeV, the three-body $\nu_4\to \nu_le^+e^-$ decay mode is allowed, and for masses above the muon mass many more decay modes open up, including 
\begin{eqnarray}
& \nu_4\to\nu_l\ell_{\alpha}^{\mp}\ell_{\beta}^{\pm}, \\
& \nu_4\to \nu_l\pi^0, \\
& \nu_4\to\ell_{\alpha}^{\mp} \pi^{\pm}, \\
& \nu_4\to\ell^{\mp}W^{\pm}, \\
& \nu_4\to\nu_l Z^0, \\
& \nu_4\to\nu_l H^0,
\end{eqnarray}
where $\ell_{\alpha},\ell_{\beta}$ are charged leptons and $H^0$ is the Higgs boson. 

Regardless of their masses, the discovery of heavy neutrinos would modify our understanding of particle physics. It would also invite several questions, including whether these new neutral particles are massive Majorana or Dirac fermions. 

As with the active neutrinos, one way to probe whether the heavy neutrinos are Majorana or Dirac fermions is to test whether they are charged under lepton number or, analogously, whether they mediate lepton-number violating processes. Heavy Majorana neutrino exchange could, for example, contribute to a nonzero rate for neutrinoless double-beta decay. Their contribution, of course, would be entangled with that of the light neutrinos and could lead to a significantly enhanced or suppressed rate relative to what is expected from light neutrino exchange. This contribution, however, is rather indirect.

One can also investigate whether the decay of the heavy neutrinos violates lepton number. For example, If the neutrino is produced in a charged-meson decay together with a charged-lepton and later decays into another charged lepton (e.g. $K^+\to\mu^+\nu_4$ followed by $\nu_4\to e^{\pm}\pi^{\mp}$), it may be straightforward to spot lepton-number violating effects. Indeed, same-sign dilepton events in a hadron collider are among the different clean search channels for Majorana neutral heavy leptons (e.g. $pp\to W^+\to\mu^+\nu_4$ followed by $\nu_4\to e^+$~plus jets). This strategy, however, may fail in a variety of ways. If the heavy neutrino is too light, it may be forbidden from decaying into a final state that easily reveals its lepton number. For example, if the heavy neutrino mass is below the pion mass, all information regarding the would-be lepton number of the final-state is contained in neutrinos, which we assume are not observable. This includes the three-body decays $\nu_4\to\nu_le^+e^-$ and $\nu_4\to\nu_l\mu^{\pm}e^{\mp}$. It is also possible that the detector cannot tell positively from negatively charged leptons. This is the case of many neutrino detectors associated with the current and the next generation of neutrino oscillation experiments (Super- \cite{Abe:2011ks} and Hyper-Kamiokande \cite{Abe:2014oxa}, the short-baseline detectors at Fermilab \cite{Antonello:2015lea}, NO$\nu$A \cite{Ayres:2004js}, DUNE \cite{Acciarri:2015uup}). 

The angular distribution of the daughters of the heavy neutrino decay, discussed in the previous section, provides another handle on revealing the nature of neutrinos, including the heavy one. In order to pursue this avenue, one needs to meet several requirements. We discuss some of these in more detail. 

\emph{The heavy neutrino sample must be polarized.} Assuming these are produced via the weak interactions, as discussed above, this is almost always the case given the maximally-parity-violating nature of the weak interactions.\footnote{One intriguing exception is when the heavy neutrinos are produced via the two-body-final-state decay of spinless charged mesons (e.g. $K^+\to\mu^+\nu_4$) and the neutrino mass equals that of the charged lepton. In $K^+ \ra \mu^+ \nu_4$, for example, the spinlessness of the parent meson requires that the $\mu^+$ and $\nu_4$ emerge with helicities of the same sign, even though the weak interactions would normally produce a right-handed $\mu^+$ and a left-handed $\nu_4$. If the $\nu_4$ and the $\mu^+$ have equal mass, then the probability for the $\nu_4$ to emerge with disfavored right-handed helicity equals that for the $\mu^+$ to emerge with disfavored left-handed helicity. Thus, there will be as many decays with two right-handed daughter leptons as with two left-handed ones. That is, both the $\nu_4$ and the $\mu^+$ will be unpolarized.} Furthermore, if neutrinos are Dirac fermions, the polarization of a produced sample of $\bar{\nu}_4$ particles will typically be opposite to that of the corresponding sample of $\nu_4$ particles. Since in the two-body $\nu_4$ and $\bar{\nu}_4$ decays $\bar{\alpha} = - \alpha$, \Eq{9}, the anisotropies with respect to a fixed direction in the $\nu_4$ and $\bar{\nu}_4$ decays will then be of the {\em same} sign. Thus, even in an experiment that cannot tag the lepton number of each neutrino, and hence can only study the sum of the $\nu_4$ and $\bar{\nu}_4$ decays, the anisotropies in these two decays will not cancel each other. Indeed, even if these anisotropies were of opposite sign, they would very likely still not cancel each other, because any accelerator-laboratory neutrino beam is produced by shooting protons at a fixed target (a charge-asymmetric initial state), and the parent mesons whose decays yield the neutrinos are often charge-selected before they decay. Hence, there will be a different number of $\nu_4$ particles than of $\bar{\nu}_4$ particles in the beam, and consequently a different number of $\nu_4$ decays than of $\bar{\nu}_4$ decays.

\begin{figure}[htb]
\includegraphics[width=5in]{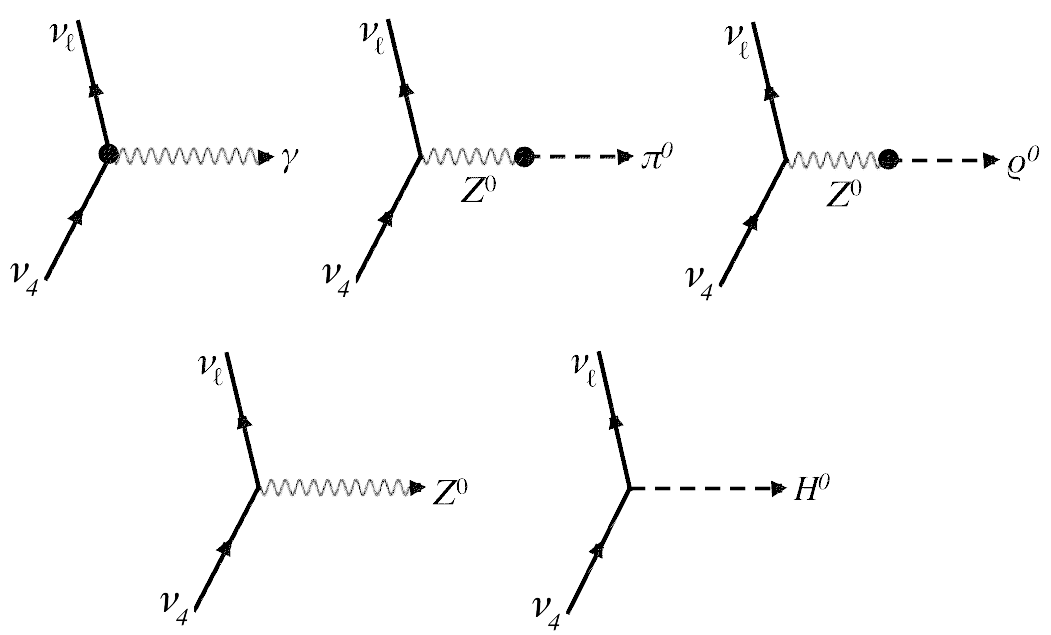}
\caption{The processes assumed to dominate, in the Dirac case, the decays $\nu_4 \ra \nu_l + X$ when $X =  \gamma,\, \pi^0,\, \rho^0, \,Z^0$, or $H^0$.}
\label{figD}
\end{figure}
\emph{The angular distribution of the $\nu_4$ decay products must be anisotropic in the Dirac case}. Figure~\ref{figD} depicts the Feynman diagrams we assumed for each of the decay modes we have considered. In Fig.~\ref{figD}, the coupling of the $\gamma$ to the neutral leptons is assumed to be via a transition magnetic dipole moment $\mu$ and an electric dipole moment $d$, while the coupling of the Higgs boson $H^0$ to the neutral leptons is assumed to be via a Yukawa interaction.

A simple illustration of the non-isotropic angular distributions of Dirac neutrino decays is provided by the decay $\nu_4 \ra \nu_l + \pi^0$. Owing to the chiral structure of the Standard Model neutral weak current, if $\nu_4$ and $\nu_l$ are Dirac particles, the amplitude for $\nu_4 \ra \nu_l + \pi^0$ is proportional to 
\beq
\bar{u}_{\nu_l} p\!\!\!/_\pi \frac{(1-\gamma_5)}{2} u_{\nu_4} = m_4 \left[ \frac{(1-\gamma_5)}{2} u_{\nu_l} \right] ^\dagger \gamma_0 \,u_{\nu_4} \;\; .
\label{eq12}
\eeq
Here, $u_{\nu_l}$ and $u_{\nu_4}$ are Dirac wave functions, $p_\pi$ is the momentum of the $\pi^0, \; m_4$ is the mass of $\nu_4$, and we have neglected the mass of $\nu_l$. So long as the $\nu_4$ and $\pi^0$ masses are not extremely close to being equal, the daughter $\nu_l$ will be highly relativistic in the $\nu_4$ rest frame. As a result, the left-handed chiral projection operator in \Eq{12} will make the amplitude for the $\nu_l$ to have right-handed helicity negligible relative to that for it to have left-handed helicity. Thus, in essentially every decay, the parameter $\lambda \equiv \lambda_\pi - \lambda_\nu$ will be +1/2, and therefore the angular distribution of the pions from $\nu_4 \ra \nu_l + \pi^0$ will be proportional to $(1+\cos\theta)$. This is as far from isotropy as it is possible to get in the two-body decay of a spin-1/2 particle [see Eqs.~(\ref{eq1})-(\ref{eq4})]. 

We have computed at leading order the decay-asymmetry parameters $\alpha$ for different neutrino decay final-states, assuming the neutrinos are Dirac fermions. These are tabulated in Table~\ref{table:alpha} and are all, in general, nonzero. Indeed, most are, in fact, order one in magnitude, with few exceptions.
\begin{table}[ht]
\caption{Decay asymmetry parameters $\alpha$ for the two-body-final-state decays $\nu_4\to\nu_l+$~boson, as defined in Eq.~(\ref{eq4}), assuming that neutrinos are Dirac fermions, $m_4$, $m_{\rho}$ and $m_Z$ are the mass of the heavy neutrino, the neutral $\rho$-meson, and the $Z$-boson, respectively. $\mu$ and $d$ are the magnetic and electric transition dipole moments. Both are generated at one-loop assuming the heavy neutrinos interact as prescribed by the weak interactions.
\vspace{4pt} }
\centering
\scalebox{1.4}{
\begin{tabular}{|c||c|c|c|c|c|}
\hline
\small{Boson}  & $\gamma$ &  $\pi^0$ & $\rho^0$ & $Z^0$ & $H^0$ \\ 
\hline  & & & & & \\ [-1em]
$\alpha$ & $\frac{2\Im(\mu d^*)}{|\mu|^2+|d|^2}$ & 1 & $\frac{m_4^2-2m_{\rho}^2}{m_4^2+2m_{\rho}^2}$ & $\frac{m_4^2-2m_{Z}^2}{m_4^2+2m_{Z}^2}$ & 1 \\ [+1em]
\hline
\end{tabular}}
\label{table:alpha}
\end{table}

It is interesting that the asymmetry parameter for the $\nu_4$ decay into a vector boson V,  $\nu_4\to\nu_l+V$, vanishes for $m_4= \sqrt{2} m_V$, where $m_V$ is the mass of the vector boson. Furthermore, depending on the relative magnitude of $m_V$ and $m_4$, it can have either sign. This is easy to understand. Angular momentum conservation allows one to write
\begin{equation}
\frac{dN (\nu_4 \to \nu_l +V)}{d \cos \theta_V} = \frac{1}{2} \Gamma_{\lambda_V = 0} (1+ \cos \theta_V) + \frac{1}{2} \Gamma_{\lambda_V = -1} 
(1- \cos \theta_V)~.
\end{equation}
The relevant amplitudes are proportional to dot products involving the polarization four-vectors for $V$ so
\begin{equation}
\label{amp0}
{\rm Amplitude} (\lambda_V = 0) \propto \frac{m_4}{m_V} 
\end{equation}
and 
\begin{equation}
\label{amp-1}
{\rm Amplitude} (\lambda_V = -1) \propto \sqrt{2}
\end{equation}
where the proportionality factors (e.g. interaction strength) are the same for both amplitudes. Using Eqs. (\ref{amp0}) and (\ref{amp-1}),
\begin{equation}
\label{asymmetryparam}
\alpha = \frac{\Gamma_{\lambda_V=0} - \Gamma_{\lambda_V = -1}}{\Gamma_{\lambda_V=0} + \Gamma_{\lambda_V = -1}} =
\frac{m_4^2 - 2 m_V^2}{m_4^2 + 2 m_V^2}. 
\end{equation}

\subsection{An Advantageous Class of Decay Modes}

Until this point we have focused on the decay modes $\nu_4 \ra \nu_\ell + X$, where $X = \bar{X}$. The theoretical discussion of these modes is particularly clean. However, with a neutrino-facility detector that can identify $e,\; \mu$, and $\pi$, but has no electric charge discrimination, one can also determine whether the neutrinos, such as $\nu_4$, are Majorana or Dirac particles by studying the decay modes 
\beq
\nu_4 \ra \ell_\alpha^\mp + X^\pm \;\; ,
\label{eqIII.14}
\eeq
where $\ell_\alpha$ is an $e$ or $\mu$, and $X$ is, for example, a $\pi$ or $\rho$. If the detector can measure the $\ell_\alpha$ and $X$ momenta, it can determine, in each event, where in momentum space the $\nu_4$ rest frame is. Thus, it can determine the angular distribution of the daughter $X$ in the $\nu_4$ rest frame directly. In addition, a peak at the $\nu_4$ mass in the $\l_\alpha X$ invariant mass distribution would help to reduce backgrounds. Consequently, the decay modes $\nu_4 \ra \ell_\alpha^\mp + X^\pm$ would appear to be experimentally advantageous.

To be sure, for given $\ell_\alpha$ and $X$, a detector that lacks electric charge discrimination can measure only the sum of the $\ell_\alpha^- X^+$ and $\ell_\alpha^+ X^-$ angular distributions. However, through an analysis similar to that which we carried out for $\nu_4 \ra \nu_\ell + X$ in Sec. \ref{sec:decay}, we find that this sum of angular distributions will be isotropic if $\nu_4$ is a Majorana fermion, but not isotropic if $\nu_4$ is a Dirac fermion. In the latter case, the detector is summing indiscriminately over $\nu_4 \ra \ell_\alpha^- + X^+$ and $\bar{\nu}_4 \ra \ell_\alpha^+ + X^-$ events.

We conclude that the angular distributions in $\optbar{\nu_4} \ra \ell_\alpha^\mp + X^\pm$, as seen in a detector that does not have charge discrimination, can also be used to determine whether neutrinos are Majorana or Dirac particles.

As an aside, let us recall that in our discussion of the decays $\nu_4 \ra \nu_l + X$, where $X = \bar{X}$, our working assumption has been that all neutrinos have the same character: either all of them are Majorana fermions, or all of them are Dirac fermions. However, an analysis similar to those already discussed shows that, regardless of the character of the daughter neutrino in these decays, the angular distribution (and the corresponding lab-frame $X$ particle energy distribution) will depend only on the character of the parent. The angular distribution will be isotropic if the parent is a Majorana fermion, and not isotropic if it is a Dirac fermion. This conclusion takes into account the fact that no realistic detector will detect the daughter neutrino. Thus, just as a detector that lacks charge discrimination can observe only the sum of the $\ell_\alpha^- X^+$ and $\ell_\alpha^+ X^-$ angular distributions in the decays $\optbar{\nu_4} \ra \ell_\alpha^\mp + X^\pm$, so, if $\nu_l \neq \bar{\nu}_l$, any realistic detector can observe only the sum of the $\nu_l X$ and $\bar{\nu}_l X$ angular distributions in the decays $\optbar{\nu_4} \ra \optbar{\nu_l} + X$.

\subsection{Other Practical Concerns}

In order to observe the decay of the heavy neutrino into a particular final state, it is imperative that the rate of decay into this final state be great enough. This, in turn, requires that the heavy neutrino lifetime be short enough. We have estimated the heavy neutrino lifetime as a function of its mass \cite{Gorbunov:2007ak,Ballett:2016opr,Berryman:2017twh}. The lifetime depends on the unknown new mixing parameters ($|U_{\alpha4}|^2$). These are constrained by existing data and the current upper bounds are strongly dependent on $m_4$~\cite{Smirnov:2006bu,Atre:2009rg,Drewes:2015iva,deGouvea:2015euy,Alekhin:2015byh,Fernandez-Martinez:2016lgt,Adhikari:2016bei,Abada:2016plb,Cai:2017mow,Drewes:2018gkc}. Roughly, in the absence of new interactions, $c\tau_4\gtrsim 10^9$~m for $m_4=10$~MeV, $c\tau_4\gtrsim 10^3$~m for $m_4=100~$MeV, and $c\tau_4\gtrsim 10$~cm for $m_4=500~$MeV.\footnote{These estimates are obtained assuming the square-magnitude of relevant elements of the mixing matrix $|U_{\alpha4}|^2\sim 0.1$, which we assume is a loose upper bound on these new mixing parameters. The lifetime, of course, scales like $1/|U_{\alpha4}|^2$. } It is safe to conclude that a significant number of heavy neutrino decays requires masses larger than tens of MeV. 

The mass region between tens of MeV and a few GeV is, not by chance, the main target of the NA48/2 and NA62 experiments at the CERN SPS \cite{Cenci:2018nlo}, and the SHiP \cite{Anelli:2015pba} and DUNE \cite{Acciarri:2015uup} proposals. These experiments can look for heavy neutrinos by producing them in meson-decay processes and observing their decays inside a large decay volume. Given current constraints on heavy neutrinos, both the DUNE and  SHiP proposals, for example, are capable of observing hundreds of heavy neutrino decays assuming these neutrinos have masses around 500~MeV. For these masses the dominant decay modes of the heavy neutrino are $\pi^0\nu_l$, $\pi^+e^-$, and $\pi^+\mu^-$. 

As alluded to earlier, in order to establish whether the heavy neutrino decay is isotropic in the rest frame of the parent neutrino, the initial state of the neutrino---i.e., its momentum---needs to be well characterized. This is especially challenging if we are interested in the decay of a heavy neutrino into a light neutrino and another Standard Model particle. Since the final-state neutrino is not measured, it is, in general, very hard to reconstruct the momentum of the parent heavy neutrino on an event-by-event basis. This issue can be bypassed, in principle, in a few ways. For example, it may be possible to learn about the kinematical properties of the heavy neutrino from its production.  In case of heavy neutrinos produced by meson decays, for example, the neutrinos inherit the momentum distribution from the parent-mesons.\footnote{This is analogous to the conditions in long-baseline neutrino experiments. There, the neutrino energies are not known on an event-by-event basis, but the neutrino energy distribution is known with some precision. Furthermore, the neutrino energy cannot be trivially measured after it scatters in the near or far detector.} This is especially convenient in decay-at-rest-beams, where the parent meson is stopped before it decays. In this case, if the heavy neutrino is the product of a two-body decay, it is monochromatic and its energy is known exactly. On the other hand, the heavy neutrino ``beam'' is isotropic and one may need to worry about the reconstruction (or lack thereof) of the direction of the heavy neutrino momentum. 

If $\nu_4$ is heavy enough to decay into $e \pi$, or perhaps even into $\mu \pi$, then, as we have discussed, its Majorana or Dirac character can be determined even in a neutrino-beam or meson-factory experiment employing detectors without charge discrimination. In addition, if, for example, we have in such an experiment a $\nu_4$ beam of known direction, and both  the $\pi^0\nu_l$ and $\pi^+e^-$ decay modes are observed, we can use the visible final state ($\pi^+e^-$) to reconstruct the energy distribution of the heavy neutrinos and then use this distribution in order to determine whether the $\pi^0\nu_l$ decay is isotropic. This could help confirm the conclusion concerning whether $\nu_4$ is a Majorana or Dirac particle drawn from study of the $e \pi$ final state alone. One last possibility is that there may be two heavy neutrinos, $\nu_4$ and $\nu_5$. In this case, one can hope to observe, for example, $\nu_5\to\nu_4\rho^0$, followed by $\nu_4\to\pi^+e^-$ and fully reconstruct the momentum of the initial-state $\nu_5$.

The mass region above tens of GeV is accessible to high-energy collider experiments, including those at the LHC \cite{Aad:2015xaa,Sirunyan:2018xiv,Cai:2017mow}. The situation here is qualitatively different. The lepton number of the initial state---zero---is well known and there are circumstances where the lepton number of the final state can be characterized well. Since these $\nu_4$s are heavy, event topologies similar to $pp\to W^+ +{\rm stuff} \to \ell^+\nu_4+ {\rm stuff}\to \ell^+\ell^{\prime +}+ {\rm stuff}$, where `stuff' stands for reconstructed particles with zero lepton number, would unambiguously reveal that $\nu_4$ is a Majorana fermion. On the other hand, knowledge of the properties of the heavy neutrinos would be available and it would be, in principle, possible to reconstruct the rest frame of the heavy neutral lepton and measure the decay angular distribution of its daughter boson $X$ in $\nu_4\to \nu_l+X$ decays. The importance of looking at angular distributions at collider experiments was also highlighted in Ref.~\cite{Han:2012vk}. It should be noted that if $\nu_4$ is heavy enough, $X=Z^0$ and $X=H^0$ may be accessible. The Higgs-final-state includes information on the neutrino Yukawa couplings.

\setcounter{footnote}{0}
\setcounter{equation}{0}
\section{Discussion and Conclusions}
\label{sec:conclusions}

Massive neutrinos are either Majorana or Dirac fermions. Given all neutrino-related information available, these two qualitatively different hypotheses are both still allowed. The reason for this is that, in the laboratory reference frame, neutrinos are almost always ultra-relativistic and, it turns out, it is very difficult to distinguish Majorana from Dirac neutrinos under these conditions. Experiments with non-relativistic neutrinos, on the other hand, have no difficulty distinguishing Majorana from Dirac neutrinos. 

Here we explored the physics of neutrino decay, concentrating on how it can be used to establish the nature of the neutrinos. Decaying neutrinos are, in some sense, always non-relativistic---you can naturally describe the decay process in their rest frame---and we anticipate that Majorana and Dirac fermions can be qualitatively different. We showed that the angular distribution of the final state boson in the two-body decay $N\to \nu_l+X$ of a polarized neutrino $N$ into a lighter neutrino $\nu_\ell$ and a self-conjugate boson $X$ is isotropic in the parent's rest frame if neutrinos, including $N$, are Majorana fermions. In contrast, if neutrinos are Dirac fermions, the angular distribution in such decays is almost never isotropic. This is a very general --- albeit approximate --- result. It depends only on CPT-invariance and is exact at leading order. It is also exact to all orders if CP-invariance is respected in the neutrino sector. 

We pointed out that while measuring the angular distribution of $X$ in the parent neutrino rest frame may be very challenging, the same information is captured, in the laboratory frame, by the energy distribution of $X$, an observable that is, perhaps, more accessible, even to neutrino-beam experiments. We identified qualitative conditions that need to be met in order to attempt such measurements. We explained that the angular distributions in two-body decays of a heavy neutrino into \emph{charged} daughter particles can also reveal whether neutrinos are Majorana or Dirac particles, even when these decays are studied by a detector that does not have charge discrimination.     

We did not explore other neutrino decay modes, including three-body final states (e.g. $N\to \nu_l\ell^+\ell^{\prime -}$). We expect these also contain robust  information capable of distinguishing Majorana from Dirac neutrinos. We hope to return to this topic in another manuscript. We also did not explore the application of this procedure to resolving the Majorana versus Dirac nature of other hypothetical particles, including the gauginos in supersymmetric versions of the Standard Model. This question has been discussed in the literature---see for example, \cite{Choi:2010gc} for a comprehensive discussion. The observation of the decay $\chi_2^0\to \chi_1^0 Z^0$, followed by the measurement of the helicity of the daughter $Z$-boson, for example, was explored in \cite{Choi:2003fs,Kim:2007zzm} as a means to address the nature of the neutralinos $\chi_2^0, \chi_1^0$.

We left out several possible sources of non-relativistic neutrinos from our discussion. The most prominent among them is the cosmic neutrino background. To detect these background neutrinos experimentally it may be possible to compensate for their very low energies using targets with vanishing threshold energies, such as beta-decaying nuclei \cite{Cocco:2007za}. The capture cross section of such neutrinos is inversely proportional to the neutrino velocity, as cross sections of exothermic reactions of non-relativistic particles typically are. In such cases, the number of capture events converges to a constant value as the velocity goes to zero, making experimental investigations somewhat more realistic 
\cite{Lazauskas:2007da}. Another possible source of non-relativistic neutrinos was suggested following the observation of a monochromatic, 3.5 keV emission line in the X-ray spectrum of galaxy clusters. Such a line may result from the decay of a 7 keV neutral fermion that decays into a photon and an active neutrino. Such neutral fermions are candidates for dark matter particles as they can be resonantly produced in the Early Universe \cite{Abazajian:2017tcc}. Other conventional non-relativistic neutrino sources have been explored in the literature \cite{Berryman:2018qxn,Yoshimura:2011ri,Yoshimura:2006nd,Fukumi:2012rn,Dinh:2012qb,Song:2015xaa}. While intriguing, the rates for the low-energy processes involving these sources are way outside the reach of even the most ambitious laboratories.

\section*{Acknowledgements}
This work was supported in part by the US National Science Foundation (NSF) Grant No. PHY-1806368 at the University of Wisconsin, in part by the US Department of Energy (DOE) grant \#de-sc0010143 at Northwestern University, and in part by the NSF grant PHY-1630782 at both universities. The document was prepared using the resources of the Fermi National Accelerator Laboratory (Fermilab), a DOE, Office of Science, HEP User Facility. Fermilab is managed by Fermi Research Alliance, LLC (FRA), acting under Contract No. DE-AC02-07CH11359.


\end{document}